# High temperature surface state in Kondo insulator $U_3Bi_4Ni_3$

Christopher Broyles,[1] Xiaohan Wan,[2] Wenting Cheng,[2] Dingsong Wu,[3]
Hengxin Tan,[4] Qiaozhi Xu,[1] Hasan Siddiquee,[1] Wanyue Lin,[1] Yuchen Wu,[1]
Jieyi Liu,[5] Yulin Chen,[3,6] Binghai Yan,[4,*] Kai Sun,[2,†] and Sheng Ran[1,‡]

[1]*Department of Physics, Washington University in St. Louis, St. Louis, MO 63130, USA*

[2]*Department of Physics, University of Michigan, Ann Arbor, Michigan 48109, USA*

[3]*Department of Physics, University of Oxford, Oxford OX1 3PU, United Kingdom*

[4]*Department of Condensed Matter Physics, Weizmann Institute of Science, Rehovot 7610001, Israel*

[5]*Diamond Light Source, Didcot OX11 0DE, United Kingdom*

[6]*ShanghaiTech Laboratory for Topological Physics, Shanghai 200031, People's Republic of China*

## Abstract

The resurgence of interest in Kondo insulators has been driven by the discovery of metallic surface state, confirmation of the non-trivial band topology, and the potential observations of charge-neutral fermions. However, the small size of the bandgap restricts scientific research and device applications to temperatures below the point of resistance saturation, where surface conduction starts to dominate the electrical transport. It is critical to explore Kondo insulators with metallic surface states at elevated temperatures. Here, we address this by reporting on a Kondo insulator, $U_3Bi_4Ni_3$, where the metallic surface state operates at significantly higher temperatures than typically observed. Our thickness-dependence and inverted transport measurements reveal that a surface state emerges below 250 K and dominates transport properties below 150 K, which is well above the temperature scale of the other topological Kondo insulators. At low temperatures, the surface conductivity is about one order of magnitude higher than that of the bulk. The robustness of the surface state indicates that it is inherently protected. This discovery not only paves the way for a deeper understanding of the interplay between Kondo physics and band structure topology but also opens up new avenues for harnessing the robust strongly correlated surface conduction in practical applications.

* Corresponding author: binghai.yan@weizmann.ac.il

† Corresponding author: sunkai@umich.edu

‡ Corresponding author: rans@wustl.edu

In Kondo lattice materials, interactions between conduction electrons and localized $f$ electrons lead to the screening of localized magnetic moments, generating a spin-scattering resonance state and a hybridization gap.[1–5] If the Fermi energy lies within the hybridization gap, the system is a Kondo insulator.[6] Band inversion can further result in a topological insulator state.[7,8] Two well-known examples are $SmB_6$ and $YbB_{12}$. Theoretical studies suggest that both materials are topologically nontrivial, possessing three-dimensional insulating bulk states and metallic two-dimensional surface states.[9–14] These topological Kondo insulators are more exotic than their non-interacting counterparts, with highly robust insulating bulk[15] and possible observations of itinerant charge-neutral fermions[16–19] that greatly challenges the theoretical understanding of a Kondo insulating state.[20–28]

The metallic surface state of the topological Kondo insulators is indicated by the low-temperature plateau of the electrical resistivity measurement, where surface conduction surpasses that of the bulk.[29,30] However, due to the small size of the Kondo gap, approximately 5-10 meV for $SmB_6$[31–34] and 15 meV for $YbB_{12}$,[32,35] the resistivity saturation only occurs at very low temperatures, typically a few Kelvin. The predominance of surface conduction at such low temperatures greatly restricts both scientific investigations and device applications. Therefore, it is crucial to identify another Kondo insulator that demonstrates metallic surface states at higher temperatures.

Uranium-based Kondo lattice systems offer a promising platform to look for high temperature topological Kondo insulators. Uranium atoms provide the large spin-orbital coupling that is essential for the band inversion .[36,37] Due to the stronger hybridization between U $5f$ electrons and conduction electrons, the hybridization gap and Kondo temperature can be significantly higher than its rare earth counterparts.[38–40] In a systematic search for uranium based topological Kondo insulators, we came across a potential candidate, $U_3Bi_4Ni_3$. It crystallizes onto a cubic lattice with non-centrosymmetric symmetry in the space group I-43d (No. 220),[41] isostructual to the Ce-based Kondo insulator $Ce_3Bi_4Pt_3$ (Fig.1a). Upon cooling below $T_K = 100$ K, there is a suppression of magnetic susceptibility[41] and an increase of correlation between $5f$ and conduction electrons,[42] both are signatures of Kondo screening of the uranium local moments. The hybridization gap was measured with photoemission spectroscopy[41] and NQR $T_1^{-1}$ spin-lattice relaxation time,[42] revealing a spin gap (19 mev) that is smaller than the charge gap (72 meV). This further supports the Kondo insulating ground state of $U_3Bi_4Ni_3$. In this work, we provide conclusive evidence of metallic surface state of $U_3Bi_4Ni_3$ that emerges below 250 K, dominating the transport properties below 150 K. This high temperature surface dominated conduction is unprecedented and could potentially enable a wide range of applications based on strongly correlated two dimensional surface conduction.

## I. KONDO INSULATING GROUND STATE

First we characterize the Kondo insulator behaviour of $U_3Bi_4Ni_3$. At high temperatures, the magnetic susceptibility, $M/H$, shows a Curie-Weiss temperature dependence, fitting to a magnetic moment of 3.68 $\mu_B$ and $T_{CW} = -162$ K (Fig. 1b). Below 100 K, a signature of Kondo screening emerges, with $M/H$ saturating to a value near $8 \cdot 10^{-3}$ emu/Oe-$U_{mol}$. This is consistent with an NMR/NQR study,[42] which demonstrated a breakdown of the scaling between the Knight shift and magnetic susceptibility below 100 K. The field dependent magnetization, $M(H)$, reveals no ferromagnetic magnetic ordering or field induced magnetic transition, with linear scaling for 5 K $\leq T \leq$ 300 K (Fig.1c).

The longitudinal resistivity ($\rho_{xx}$) resembles that of a narrow gap semiconductor at room temperature, with $\Delta = 95$ meV (Fig.1d). This gap size is significantly larger than those observed in $SmB_6$ and $YbB_{12}$.[31–35] As the temperature is lowered past the Kondo temperature ($T_K \approx 100$ K), $\rho_{xx}$ flattens around 185 mΩ-cm, before an additional increase below 50 K, leading to a saturation at 290 m$\Omega$-cm. An Arrhenius fit for 30 K $\leq T \leq$ 60 K gives an activation gap of $\Delta = 1.5$ meV. This temperature dependence could be consistent with Kondo hybridization renormalizing and reducing $\Delta$ below 100 K; however, the low temperature saturation in $\rho_{xx}$ is more than an order of magnitude larger than previously reported.[41]

The magneto-transport measurements, displayed in Figure 1e-g, provide further support of the Kondo coherent ground-state. The high temperature Hall resistivity ($\rho_{xy}$) exhibits a sigmoid-like shape, which can not be accurately represented by a simple two-band model and is typically indicative of the anomalous Hall effect (AHE).[43] The anomalous Hall is most clearly seen at 150 K, with a distinct, sharp jump near zero magnetic field. By incorporating a constant into the two band Hall model,[44] $\rho_{xy}^{AHE}$ is extracted and the temperature dependence is plotted in Figure 3h. The anomalous Hall contribution gradually increases from room temperature to 150 K, after which it begins to decrease. This overall temperature dependence of the anomalous Hall effect is consistent with what has been observed in other Kondo lattice systems.[45] Above the Kondo temperature, incoherent Kondo scattering dominants the transport properties. A modest magnetic field can polarize the local moments, leading to the anomalous Hall effect through a scattering mechanism.[45,46] As the temperature decreases, a smaller magnetic field is needed to polarize the local moments; thus, the curve narrows and transitions into a sharp, step-like shape at 150 K. In the meantime, Kondo scattering logarithmically increases with the decreasing temperature, enhancing the anomalous Hall effect.[47,48] Once Kondo scattering becomes coherent below $T_K$, the anomalous Hall gradually decreases. Note that below 50 K, the anomalous Hall contribution becomes negligible and the Hall data fits equally well with a pure two-band model.

To further confirm the Kondo ground state of $U_3Bi_4Ni_3$, we performed angle-resolved photoemission spectroscopy (ARPES) measurements at $T = 20$ K . To highlight the spectral profile of U $5f$ character, we used 98 eV photon energy, which resonantly enhances the photoionization cross section of U $5f$ states. Figure 2 presents the energy spectra for the resonant condition at 98 eV, and the off-resonant condition at 92 eV for comparison. While both spectra show a strong maximum at -2 eV, indicative of a flat band, the resonant enhanced band image maintains a strong intensity until $E_F$. The integrated energy distribution curve (EDC) reveals a refined structure near the Fermi level, with two peaks within 300 meV of $E_F$ (indicated by red arrows). These weak peaks indicate U $5f$ multiplet states, which participate in the many-bondy Kondo resonance, as seen in other uranium-based Kondo lattice systems.[49,50] The EDC of both 98 and 92 eV is suppressed at Fermi energy, indicating the insulating nature. The suppressed overall intensity at the Fermi level and the enhanced U $5f$ spectral weight near the Fermi energy in the resonant spectrum are consistent with a Kondo insulator, similar to observations in $SmB_6$ and $YbB_{12}$.[51–53]

## II. METALLIC SURFACE STATE

The temperature dependent resistivity of $U_3Bi_4Ni_3$ exhibits a plateau below 150 K, shown in Fig. 1d. Similar plateaus have been observed in both $SmB_6$[29] and $YbB_{12}$,[30] indicative of either a metallic surface state or extended in-gap states. To identify the origin of the plateau, we performed a thickness dependant study. The sample was successively polished to 150 $\mu$m and 120 $\mu$m. Above 200 K, the resistivity is essentially independent of thickness, which is a clear indication of bulk transport (Fig. 1d). In contrast, at low temperatures the resistivity markedly decreases as the thickness is reduced. This reduction suggests that the resistance no longer scales linearly with thickness, a clear sign of surface charge transport.

In order to more accurately determine the bulk and surface transport coefficients, we employ a non-local transport measurement, with the inverted resistance setup.[54] The inverted transport measurement makes use of the Corbino-disk geometry depicted in Figure 3a, where electrode 2 acts as a two-dimensional Faraday cage for the surface area enclosed by this electrode. This technique has been reliably used to decouple the surface and the bulk conduction channels of $SmB_6$ and other strongly correlated insulators.[15,55] Within this setup, we can measure two distinct four-terminal resistances: $R_{12;12}$ and $R_{12;34}$, where the first pair of indices denotes the current-carrying electrodes, and the second pair represents the voltage-measuring electrodes. For bulk-dominated transport, both $R_{12;12}$ and $R_{12;34}$ are proportional to the bulk resistivity. Conversely, in the scenario where surface transport dominates, $R_{12;12}$ and $R_{12;34}$ will show distinctly different dependencies on surface and bulk resistivity. This distinction arises because, when current is introduced between electrodes (1) and (2), charge carriers traverse parallel paths from 1⟶2 and 1⟶3⟶4⟶2, leveraging the enhanced conductivity of the surface. Therefore, in cases of surface-dominant transport, $R_{12;12}$ is directly proportional to the surface resistivity ($R_{12;12} \propto \rho_s$), while $R_{12;34}$ receives mixed contributions from both bulk and surface resistivity and exhibits a scaling behavior characterized by an inverse-gap dependence, (i.e., $R_{12;34} = C\ t\frac{\rho_s^2}{\rho_b} \propto \mathrm{Exp}[-\frac{\Delta}{k_B\ T}]$),[54] where 'C' is a constant, 't' is the sample thickness, '$\rho_s$' is the surface resistivity, '$\rho_b$' is the bulk resistivity, '$\Delta$' is the energy gap, '$k_\mathrm{B}$' is the Boltzmann constant, and '$T$' is the temperature.

Multiple $U_3Bi_4Ni_3$ samples were polished to a thickness ranging between 130 and 210 $\mu$m, and the Corbino disk electrodes were patterned on top and bottom faces. Here we present two Corbino-disk samples in Figure 3, which have been measured in the $R_{12;12}$ and $R_{12;34}$ configurations. The $R_{12;12}$ configuration has similar behavior to $\rho_{xx}(T)$, but only saturates at $R(5K)/R(400K) \approx 10$ (Fig. 3b,c). When measured in the $R_{12;34}$ configuration, the peak values reaches roughly 10 before the insulator-metal crossover. To compare the activated scaling, Arrhenius fittings were applied between 200 K and 300 K, revealing a gap ranging from $78.8-82.8$ meV and $56.5-74.7$ meV for $R_{12;34}$ and $R_{12;12}$ respectively. Below the surface state onset, $R_{12;34}$ displays a dual inverse-gap behavior. The initial gap has values ranging from $\Delta^1_{\mathrm{low}} = 17.7-39.3$ meV and transitions to a much smaller gap of $\Delta^2_{\mathrm{low}} =$ 2.4-3.1 meV. At low temperatures, sample 1 (S1) has a rise in $R_{12;34}$ around 25 K. The variety of these surface transport features demonstrates the robust nature of the surface state to carrier doping and impurities that characterize the sample variation.

The use of finite element analysis (FEA) allows the separation of the bulk and surface transport properties. We conducted simulations of the charge transport using a geometry that mirrors the experimental setup. By employing the bulk resistivity ($\rho_b$) and surface conductivity ($\sigma_s$) as fitting parameters, we compare the simulated outcome with the experimental values of $R_{12;12}$ and $R_{12;34}$.

This allowed us to extract the fitted values of $\rho_b$ and $\sigma_s$ at varying temperatures. The results of these fittings are displayed in Figure 3d-i. The bulk resistivity of S1 is plotted versus $T^{-1}$ in Figure 3(d), which shows an activation gap of 97.4 meV above 150 K and a much smaller activation gap of 3.2 meV below $T_K = 100$ K. As the temperature drops below 30 K, $\rho_b$ saturates, likely due to the contribution of an impurity band or other in-gap state at low temperatures. The surface conductivity of S1 significantly increases from 250 K and exhibits nearly temperature independent behavior below 150 K. The ratio of $\sigma_b * t \ / \ \sigma_s$ reaches unity at 150 K, marking a crossover in the transport properties, which is just below the peak in $R_{12;34}$ at 180 K (Fig. 3f). At the lowest temperature, $\sigma_s$ is roughly 5 times larger than the bulk, which is significant for a bulk thickness of $t = 210$ $\mu$m.

The sample variation of the surface state is highlighted in the FEA of Corbino-disk sample 2 (S2). The low temperature bulk gap is comparable at $\Delta = 4.2$ meV, but $\sigma_b$ continues to decrease with temperature (Fig. 3g). Likewise, $\sigma_s$ keeps increasing below 120 K, rather than reaching a temperature independent value (Fig. 3h). Due to the large conductivity of the bulk, the ratio of $\sigma_b * t \ / \ \sigma_s$ does not reach unity until 50 K (Fig. 3i). However, at base temperature, surface conductivity reaches 10 times that of the bulk.

## III. DISCUSSIONS

In the surface dominant regime, each Corbino device has different behavior at low temperatures. While the inverted resistance of sample 1 does not continue to decrease at low temperatures, sample 2 continues with a steep decrease until the 5 K. Similar behavior has been seen in $SmB_6$, where the non-stoichiometrically grown samples exhibit flattening of the inverted resistance at low temperature,[15] attributed to bulk impurities. FEA also reveals a difference of one order of magnitude in the bulk conductivity, likely due to the impurity bands. Despite the bulk impurities, the surface state of $U_3Bi_4Ni_3$ is dominant in a large temperature range and reaches one order of magnitude higher than its bulk conductivity.

Given that surface-dominated transport has been universally observed across all samples and surfaces investigated in our study, we conclude that this surface state is exceptionally robust and does not appear to be affected by surface conditions and impurities. This resilience is notable especially considering that the surfaces have been exposed to air and contaminants during the polishing process. Such durability is highly unusual, indicating that the surface states may be inherently protected.

Among the known mechanisms for robust surface states, topological band inversion stands out as the most plausible origin for such persistent surface conductivity. However, due to Kondo hybridization, simple DFT+U calculations may not be able capture the electronic structure of $U_3Bi_4Ni_3$ (see SI). More advanced methods, such as DMFT, are needed to fully understand the topology of $U_3Bi_4Ni_3$.

Regardless of their precise nature, the surface states observed in this study present a novel example of a robust, two-dimesional (2D) electron gas system. The interplay of Kondo physics within these 2D surface states introduces strong electron correlations and significant spin-orbit coupling, which is anticipated to exceed that of other well-studied 2D gases, such as semiconductor interfaces, graphene and transition metal dichalcogenides. Since strong interactions and spin-orbit coupling have long been recognized as crucial for the emergence of exotic states, such as fractional topological phases, it is plausible to expect that these surface states could be of profound significance for future research, irrespective of their origin.

In the prime example of the topological Kondo insulator, $SmB_6$, the onset of Kondo hybridization begins at 60-90 K and the full coherence is established below 30-40 K.[33,34] The surface state, emerging at 4 K, arises from the Kondo ground-state.[56,57] Similarly, $YbB_{12}$ has a Kondo temperature of 200 K,[35,58] but the resistivity saturation does not happens until below 2 K.[30] $U_3Bi_4Ni_3$ is on the other side the spectrum in terms of the energy scale, with the surface state appearing above the Kondo temperature: the emergence of the surface state has a concrete temperature scale between 200 - 250  K, while the Kondo energy scale is set at around $T_K = 100$ K. This difference makes $U_3Bi_4Ni_3$ a unique platform to investigate the interplay between Kondo hybridization and the metallic surface state. Since the surface state emerges at a much higher temperature than the Kondo temperature, it could be associated with a topological gap that already exists at high temperatures. Upon cooling through the Kondo temperature, the gap is renormalized and reduced in size. This is evidenced by the dual-gap features observed in both standard and inverted resistance transport, as well as in the true bulk conductivity revealed though FEA. However, in the entire temperature range, there is no gap-closure due to the renormalization, which indicates that the topology remain the same through the Kondo coherence, since a change in topology would require the closing and reopening of a gap.

The sheet conductivity of $U_3Bi_4Ni_3$ at 5 K is 0.1-0.5 $1/\Omega$, which is an order of magnitude greater than that of $SmB_6$ (0.02 $1/\Omega$)[15] and FeSi (0.01 $1/\Omega$),[55] and is comparable to that of $Bi_2Se_3$.[59] Such high conductivity indicates an unexpectedly high mobility, which is typically rare in strongly correlated systems. However, the confinement of electrons to a two-dimensional plane can modify the electronic structure and potentially change how strong correlations affect mobility, leading to different behavior compared to the bulk. Further more, if the surface state is topologically protected, it could experience reduced back scattering, potentially preserving higher mobility despite strong correlations. This high conductivity is crucial for applications requiring robust and efficient electronic transport, such as quantum computing and advanced spintronic devices.

## IV. CONCLUSION

Our results establish $U_3Bi_4Ni_3$ as a Kondo insulator that not only features a robust surface state with high sheet conductivity but also operates over an exceptionally broad temperature range. This surface state, marked by strong correlation effects, opens up numerous possibilities for innovative applications. For instance, constructing heterostructures directly on this bulk sample could exploit the strongly correlated 2D surface state, potentially leading to phenomena such as quantum Hall states or Majorana fermions when combined with superconductivity. Furthermore, if the surface state is confirmed to be topological, it could be utilized in spintronics applications, benefiting from topological protection. Given these unique opportunities, We anticipate a surge of research activities focused on exploring and utilizing this material.

## V. METHODS

**Sample Synthesis** Single crystals of $U_3Bi_4Ni_3$ were prepared using the molten metal Flux method, with Bi as the solvent. Following the reported procedure ,[41] the raw elements were prepared in molar ratios of U:Bi:Ni = 1:10:2 and placed into an Alumina crucible, which was sealed in a quartz tube with Argon gas at 0.265 atm. The furnace was heated to 1150 °C, and then cooled to a temperature of 650 °C at 5 °C/hr. To separate the crystals from the Bi Flux,the quartz tube was then taken out of the furnace at 650 °C and quickly inverted and placed into a centrifuge for 10 seconds. The $U_3Bi_4Ni_3$ single crystals were confirmed with powder X-Ray diffraction (Cu-$K_\alpha$) with a Rigaku MiniFlex.

**Sample Preparation and Measurement** The samples prepared for resistivity measurements by polishing to a thickness of $100-200$ $\mu$m and Pt wire was attached using silver epoxy. A Hall bar is prepared with a standard four-wire measurement, through a pair of Hall contacts inserted between the voltage leads. The Corbino-disk geometry is prepared by drawing a ring of photo-resist on the sample face and around the sample edge, to prevent from shorting the top and bottoms surfaces.

A gold sputter deposition of 50 nm forms an inner and outer electrode. This process is repeated on the bottom surface, to form another pair of electrodes. Electronic transport and magnetization measurements were performed in the Quantum Design Physical Property Measurement System using the Resistivity and Vibrating Sample Magnetometer options.

**Finite Element Analysis** We studied the dependence of $R_{12;12}$ and $R_{12;34}$ on bulk to surface conductivity ratio by performing finite element analysis using COMSOL MULTIPHYSICS AC/DC module. We used Electric Currents and Electric Currents in Layered Shells interfaces of the AC/DC module to build the bulk and surface conducting channels, respectively. We used the same dimensions as the experimental samples in our simulation. We fit $\frac{R_{12;12}}{R_{12;34}}$ vs bulk to surface conductivity ratio obtained from numerics with $\frac{R_{12;12}}{R_{12;34}}$ vs $T$ obtained from experiment to extract bulk to surface conductivity ratio as a function of $T$. Then we utilize the bulk to surface conductivity ratio vs $T$ data, $R_{12;12}$ (or $R_{12;34}$) vs $T$ data from experiment and $R_{12;12}\sigma_s$ (or $R_{12;34}\sigma_s$) vs bulk to surface conductivity ratio from simulation to extract bulk conductivity and surface conductivity as a function of $T$ separately.

**Angle-Resolved Photoemission Spectroscopy Measurement** ARPES measurements were performed at beamline I05 of the Diamond Light Source (DLS). The angle resolution was $\leq 0.2°$ and the overall energy resolution was set $\leq$ 25 meV. The 98 eV and 92 eV photon energy were used to examine the difference between measurements performed on- and off-resonance at the uranium O edge. The fresh surfaces for ARPES measurement were obtained by cleaving the samples in situ in the measurement chamber. During the measurements, the chamber pressure was kept below $2\times10^{-10}$ mbar, and the sample temperature was 6 K.

**Density Functional Theory** Electronic structures are calculated with the Density functional theory as implemented in the Vienna ab-initio Simulation Package,[60] combining the PBE-type generalized gradient approximation[61] for the exchange-correlation interaction between electrons. The experimental crystal structure is employed. A cutoff energy for the plane-wave basis set is 350 eV. The k-mesh grid for the Brillouin zone is 8×8×8. The Hubbard U correction for both the Ni d (Ud=0, 2, 4 eV) and U f (Uf=0, 2, 4, 6 eV) orbitals is employed. Spin-orbital coupling is included throughout. The wannier90 software[62] interface is employed to obtain the Wannier Hamiltonian for Wannier charge center calculations.

## ACKNOWLEDGMENTS

We acknowledge helpful discussions with Qimiao Si. **Funding:** Research at Washington University was supported by the National Science Foundation (NSF) Division of Materials Research Award DMR-2236528. CB acknowledges the NRT LinQ, supported by the NSF under Grant No. 2152221. Research at University of Michigan was supported by the Air Force Office of Scientific Research through the Multidisciplinary University Research Initiative, Award No. FA9550-23-1-0334 (XW and KS) and the Office of Naval Research through the Multidisciplinary University Research Initiative, Award No. N00014-20-1-2479 (WC and KS). BY acknowledges the financial support by the European Research Council (ERC Consolidator Grant “NonlinearTopo”, No. 815869) and the ISF - Personal Research Grant (No. 2932/21) and the DFG (CRC 183, A02). YLC acknowledges the support from the National Natural Science Foundation of China (No. U23A6002). We acknowledge Diamond Light Source for time on Beamline I05 under proposal number SI36513.

**Author contributions:** S. Ran, K. Sun and B. Yan conceived and designed the study. C. Broyles synthesized the single crystalline samples of $U_3Bi_4Ni_3$. C. Broyles and H. Siddiquee performed the electrical transport and magnetic measurements. G. Xu fabricated devices using focused ion beam. X. Wan, W. Cheng, W. Lin, Y. Wu and K. Sun performed the finite element analysis. D. Wu and Y. Chen performed the angle-resolved photoemission spectroscopy measurement. H. Tan and B. Yan performed the Density Functional Theory calculations. C. Broyles and S. Ran wrote the manuscript with contributions from all authors. **Competing Interests:** The authors declare no competing interests.

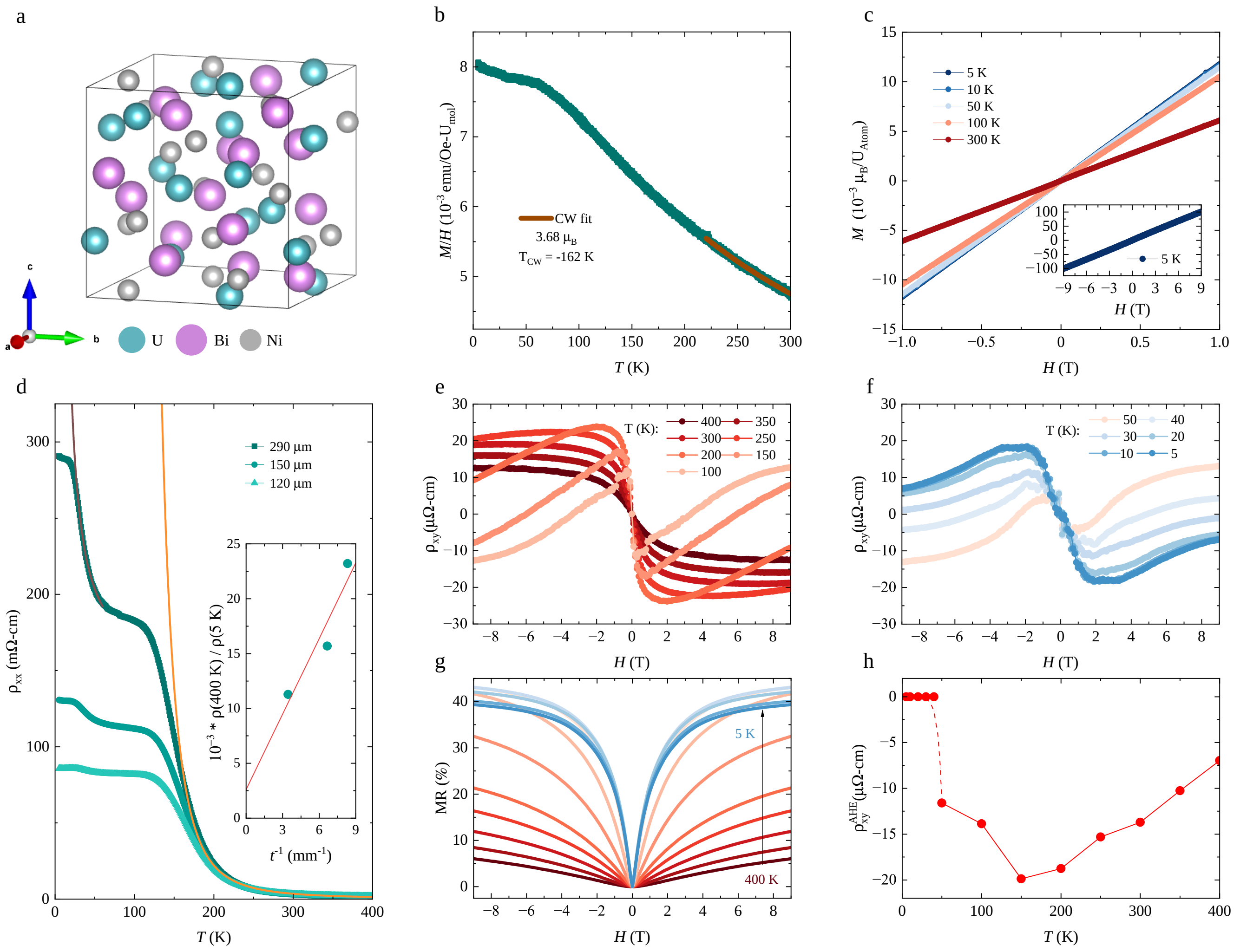

**Fig.** 1. **Bulk Kondo Insulator Characterization (a)** The crystal structure of $U_3Bi_4Ni_3$, space group No. 220 (I-43d). **(b)** The magnetic susceptibility measured from room temperature down to 5 K, with a magnetic field of 0.1 T. **(c)** Magnetic field sweeps up to 1 T for various temperatures between 5 and 300 K. The inset displays a 9 T field sweep at 5 K. **(d)** The temperature dependence of $\rho_{xx}$, measured on a single crystal which was successively polished to thicknesses of 290, 150, and 120 $\mu$m, with a fitting applied above 200 K (orange, $\Delta$=95 meV) and between 30 and 60 K (brown, $\Delta$=1.5 meV). The inset displays $\rho(400\text{ K})/\rho(5\text{ K})$ vs $t^{-1}$, modeled by the equation $\rho(400\text{ K})/\rho = (2\,\sigma_s/\sigma_{400\text{ K}})\,t^{-1} + \sigma_b/\sigma_{400\text{ K}}$. The linear fitting provides a slope of $2.3 \pm 0.8$ and an intercept of $3 \pm 5 * 10^{-3}$. **(e, f)** The Hall resistivity ($\rho_{\text{xy}}$) field sweeps for $T \geq 100$ K are displayed in (e), while those for $T \leq 50$ K are shown in (f). **(g)** The magneto-resistance measured up to 9 T and presented for a variety of temperatures between 300 K and 5 K. **(h)** The anomalous Hall resistivity ($\rho_{\text{xy}}^{\text{AHE}}$), extracted from the two-band + $C$ fittings (see SI).

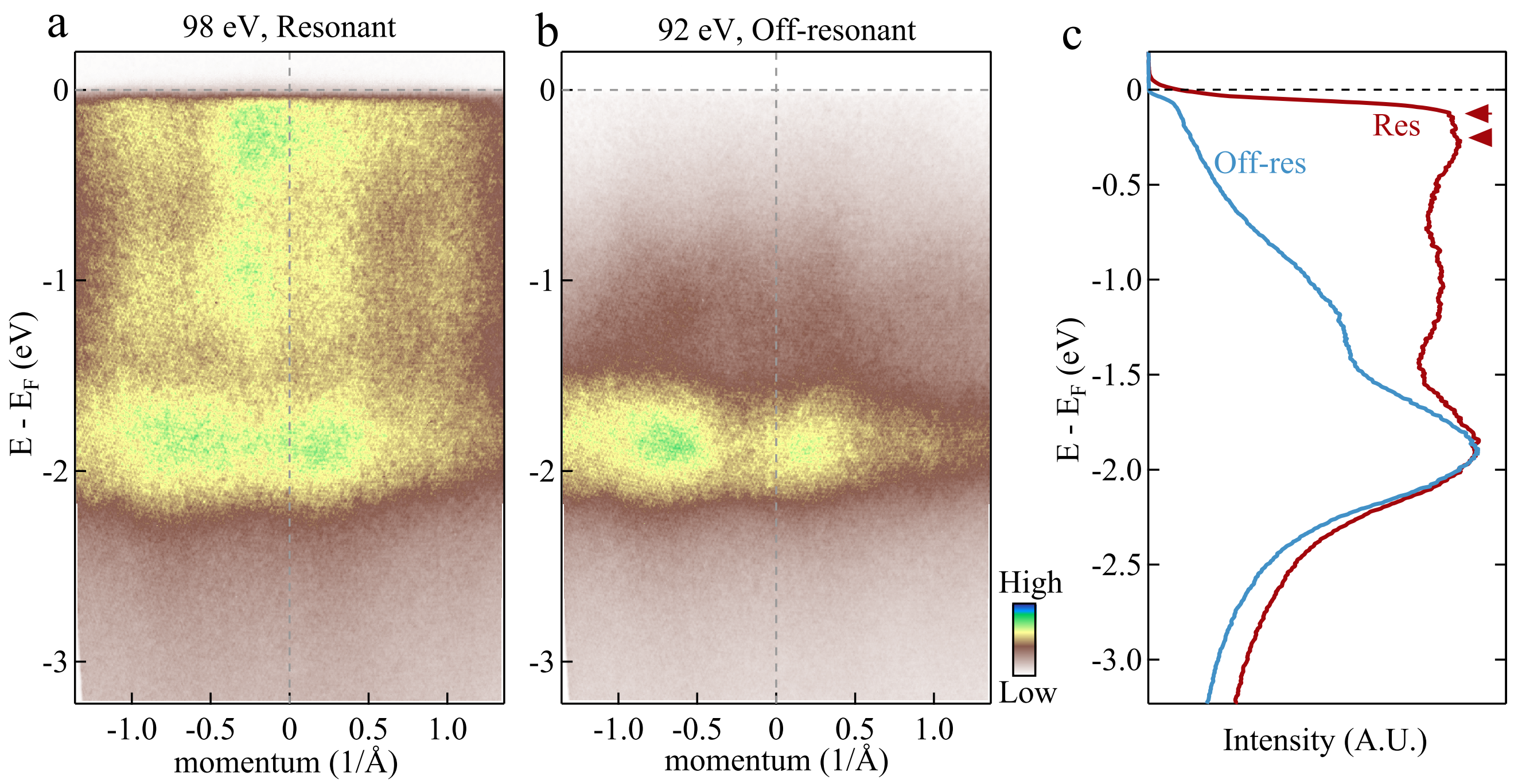

**Fig.** 2. **Resonant Enhanced ARPES (a)** The intensity map of Energy versus Angle for the resonant condition of 98 eV. **(b)** The off-resonant condition, with a photon energy of 92 eV. **(c)** The integrated energy distribution curve (EDC) of the resonant and off-resonant conditions, demonstrating $f$-band weight near $E_F$.

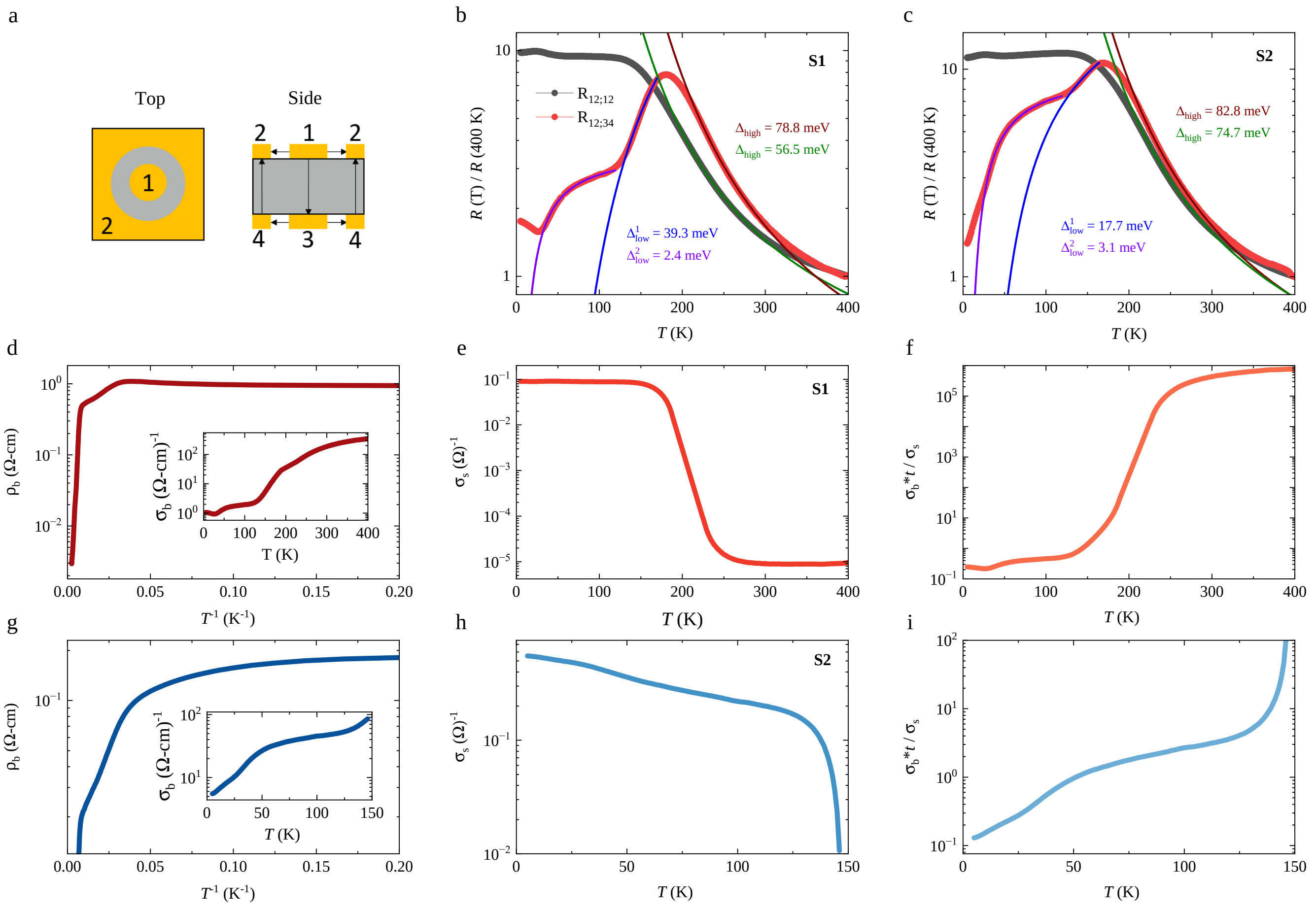

**Fig.** 3. **Corbino-disk and Finite Element Analysis (a)** The measurement geometry of the Corbino-disk, shown from the top and cross-sectional side view. **(b, c)** Corbino-disk samples 1 and 2 measured in the $R_{12;12}$ and $R_{12;34}$ configurations. **(d, e)** Using finite element analysis (FEA) to simulate the transport of sample 1, the bulk resistivity is plotted in (d), with the bulk conductivity in the inset, and the surface conductivity is displayed in (e). **(f)** The ratio of the bulk and surface conductivity, with $t = 210\ \mu$m. **(g-i)** The FEA results for sample 2, with a thickness of 130 $\mu$m, presented below the surface state onset.

# SI : High temperature surface state in Kondo insulator $U_3Bi_4Ni_3$

Christopher Broyles,[1] Xiaohan Wan,[2] Wenting Cheng,[2] Dingsong Wu,[3]
Hengxin Tan,[4] George Xu,[1] Hasan Siddiquee,[1] Jenny Lin,[1] Jack Wu,[1]
Jieyi Liu,[5] Yulin Chen,[3,6] Binghai Yan,[4,*] Kai Sun,[2,†] and Sheng Ran[1,‡]

[1]*Department of Physics, Washington University in St. Louis, St. Louis, MO 63130, USA*

[2]*Department of Physics, University of Michigan, Ann Arbor, Michigan 48109, USA*

[3]*Department of Physics, University of Oxford, Oxford OX1 3PU, United Kingdom*

[4]*Department of Condensed Matter Physics,*
*Weizmann Institute of Science, Rehovot 7610001, Israel*

[5]*Diamond Light Source, Didcot OX11 0DE, United Kingdom*

[6]*ShanghaiTech Laboratory for Topological Physics,*
*Shanghai 200031, People's Republic of China*

## I. MAGNETO TRANSPORT FITTINGS

The transverse and longitudinal resistivity were measured on single crystals of $U_3Bi_4Ni_3$, in magnetic fields up to 9 T. The symmetric component of the change in the longitudinal signal ($\Delta\rho_{xx}$) and asymmetric component of the transverse voltage ($\rho_{xy}$) are used in the fittings. The two band[1] + $C$ fittings are performed in Origin Lab, using a simultaneous fitting of $\Delta\rho_{xx}(H)$ in the range of -9 T $\leq H \leq$ 9 T and $\rho_{xy}(H)$ in a range above polarization field dependence (2 T $\leq H \leq$ 9 T). A fit sharing both the mobility and number density was unable to converge, so the number density was fit independently while the mobility was shared. Once the fitting for a field sweep was converged, the constant offset from the two band model is taken to be $\rho_{xy}^{AHE}$.

The sample fittings for the 300 K and 5 K fields sweeps are shown in SI Figure 1. The 300 K fitting shows a good agreement in the selected range and is extended to the full field range for clarity. The simultaneous fitting with $\Delta\rho_{xx}$ is shown in SI Figure 1b. The low temperature fitting for the Hall signal shows excellent agreement with the conventional two band field-dependence in SI Figure 1c, and the longitudinal component also has a qualitatively good fit. The longitudinal component does have a mismatch with the fitting endpoints. This is due to the MR not completely saturating at high fields, despite being the expectation from the Drude description.[1]

The results of the fittings are shown in SI Figure 2. In the high temperature regime, the electronic transport is dominated by the hole-type carriers. The hole number density, $\eta_h$, is at a maximum around 300 K and the activated carrier are removed as the temperature decreases. This trend continues until a crossover in the carrier at 150-200 K, which aligns with the onset of the surface state in the Corbino disk measurements. Below this temperature, $\eta_e$ overtakes $\eta_h$ by nearly 4 orders of magnitude. A similar phenomena is seen in the mobility parameters, where the $\mu_h$ surpasses $\mu_e$ by 5 orders of magnitude. Below 50 K, there is a signature of the hole-like band hybridizing with the the $f$-orbital moments, resulting in an enhanced number density with suppressed mobility. The temperature dependence can be understood through the energy scale of the surface state, shifting the chemical potential, and Kondo interaction, renormalizing the band gap.

[*] Corresponding author: binghai.yan@weizmann.ac.il
[†] Corresponding author: sunkai@umich.edu
[‡] Corresponding author: rans@wustl.edu

## II. FOCUSED ION BEAM DEVICE

A focused ion beam (FIB) device was milled and extracted from a single crystal sample of $U_3Bi_4Ni_3$, using the Thermofisher Scios 2 DualBeam FIB. The lamella of 1 $\mu$m thickness was cut into a Hall bar device, with 6 lead configuration. An image of the device, with the deposited Pt contacts, is displayed in SI Figure 3. The longitudinal and transverse resistivity is measured in zero magnetic field, from 400 K to 2 K. $\rho_{xx}(T)$ has a similar two gap scaling as the bulk measurements in the main text; but, instead of a plateau at 100 K, it appears more as a "hump-like" feature. This contrast is likely related to the Gallium doping of the FIB process, specifically impacting the crystalline environment near the surface. Further investigation is needed to fully understand the implication of defects on the electronic properties.

## III. FIRST PRINCIPLES CALCULATIONS

Results for DFT+U calculations are shown in SI Figure 4. The calculations predict a metallic band structure when Hubbard $U$ correction is not applied. The U $f$ orbitals notably contribute to the bands near the charge neutral point, while the Ni $d$ and Bi $p$ orbitals are fully occupied and form the deeper valence bands. This metallic state clearly contradict the experimental results. Consequently, we applied Hubbard $U$ corrections to both U $f$ and Ni $d$ orbitals. The Hubbard $U$ correction to the Ni $d$ orbitals has minimal impact on the band structure, because these orbitals are completely filled. Upon applying Hubbard $U$ to the uranium $f$ electrons, the U $f$ bands at the Fermi energy become split, creating a global band gap. While at moderate Hubbard $U$ value, $U = 2$ eV, both conduction and valence bands are still dominated by $f$ bands, for $U = 4$ and 6 eV, the $f$ bands are located far away from the Fermi level, and the gap forms between the dispersive Ni $d$ and Bi $p$ bands, not the U $f$ bands. This contradict the behavior expected of $U_3Bi_4Ni_3$ as a Kondo insulator, where $f$ bands are supposed to contribute to the band gap, as supported by our transport and ARPES data. Furthermore, the predicted band gap is significantly larger than the observed value. These results reflect the limitations in dealing with the Kondo lattice system that is generic to DFT+$U$ calculations, which consider the correlation effect as a local onsite interaction. In Kondo systems, the key interaction is between the localized $f$ electrons and the conduction electrons, which involves dynamic screening and hybridization processes that DFT+$U$ does not capture effectively. To better understand the electronic structure of $U_3Bi_4Ni_3$, we will need more advanced methods like DMFT.

---

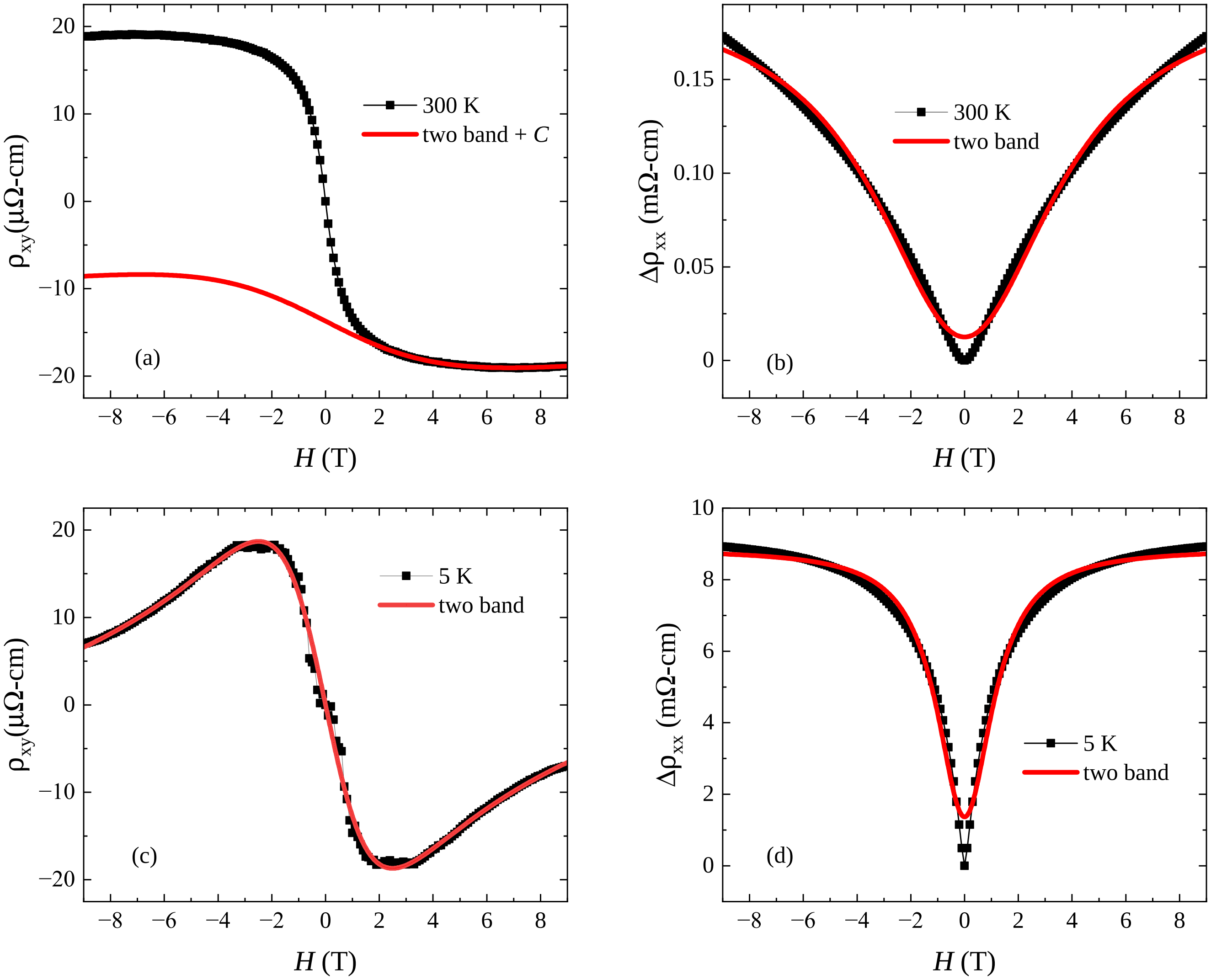

**Fig.** 1. **(a)** Sample fitting for the 300 K field sweep for the asymmetric transverse resistivity, $\rho_{xy}^{assym}$. **(b)** The change longitudinal resistance, $\Delta\rho_{xx}$, fit with the conventional two band model. **(c,d)** Sample fitting for 5 K measurement, with fitting using the whole data range and $C$ set to zero.

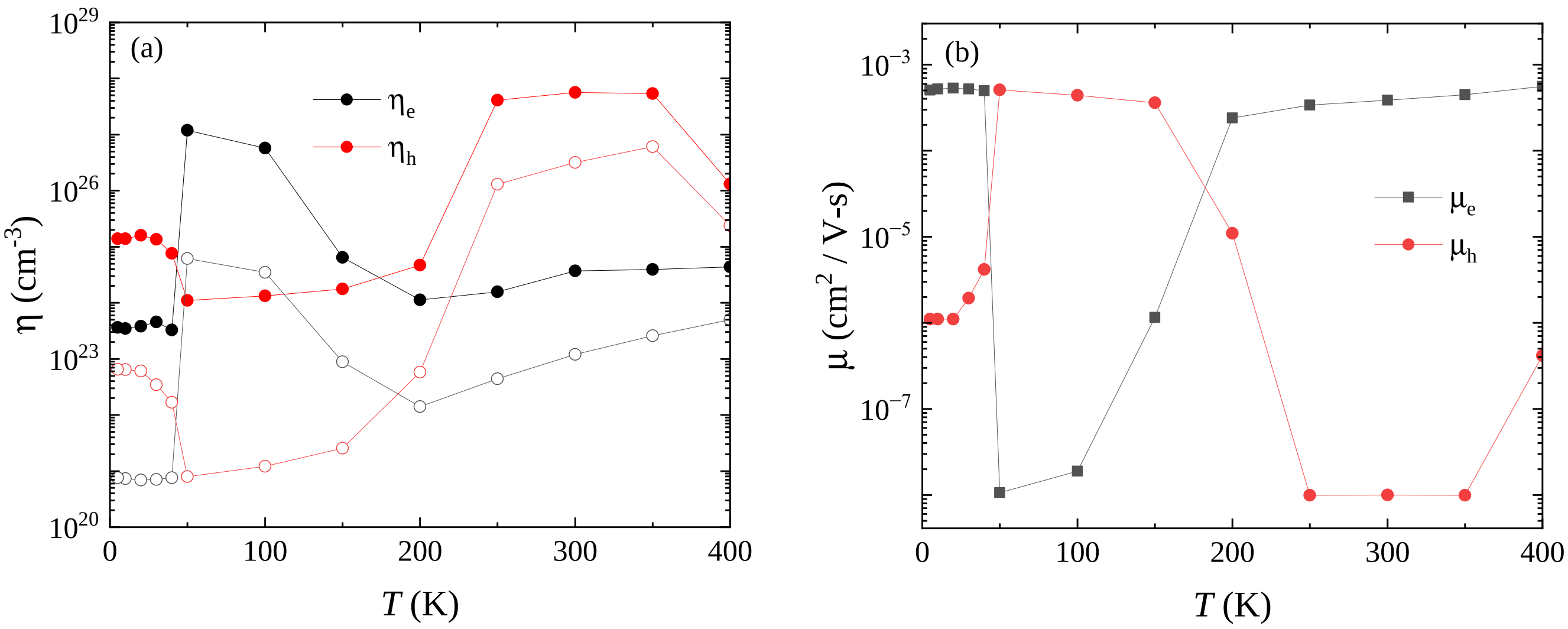

**Fig.** 2. **(a)** The number density from the two band fitting parameters, which was fit independently for $\rho_{xy}^{assym}$ (solid circles) and $\Delta\rho_{xx}$ (open circles). **(b)** The mobility parameters for electron and hole carriers, which was the shared parameter in the fittings.

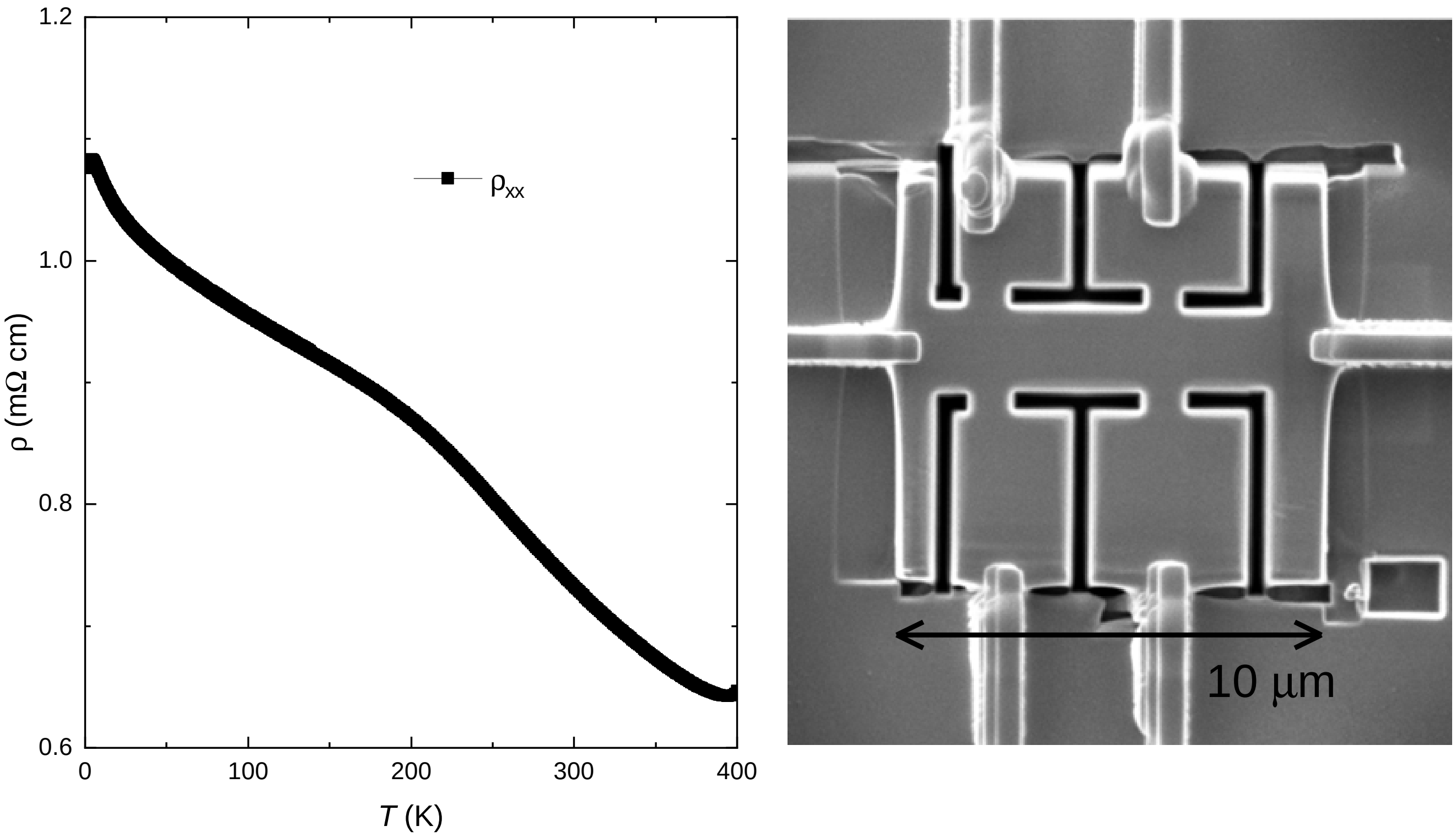

**Fig.** 3. The electronic transport characterization for the focused ion beam device, measured in zero magnetic field. An SEM image of the device, with a thickness of 1 $\mu$m.

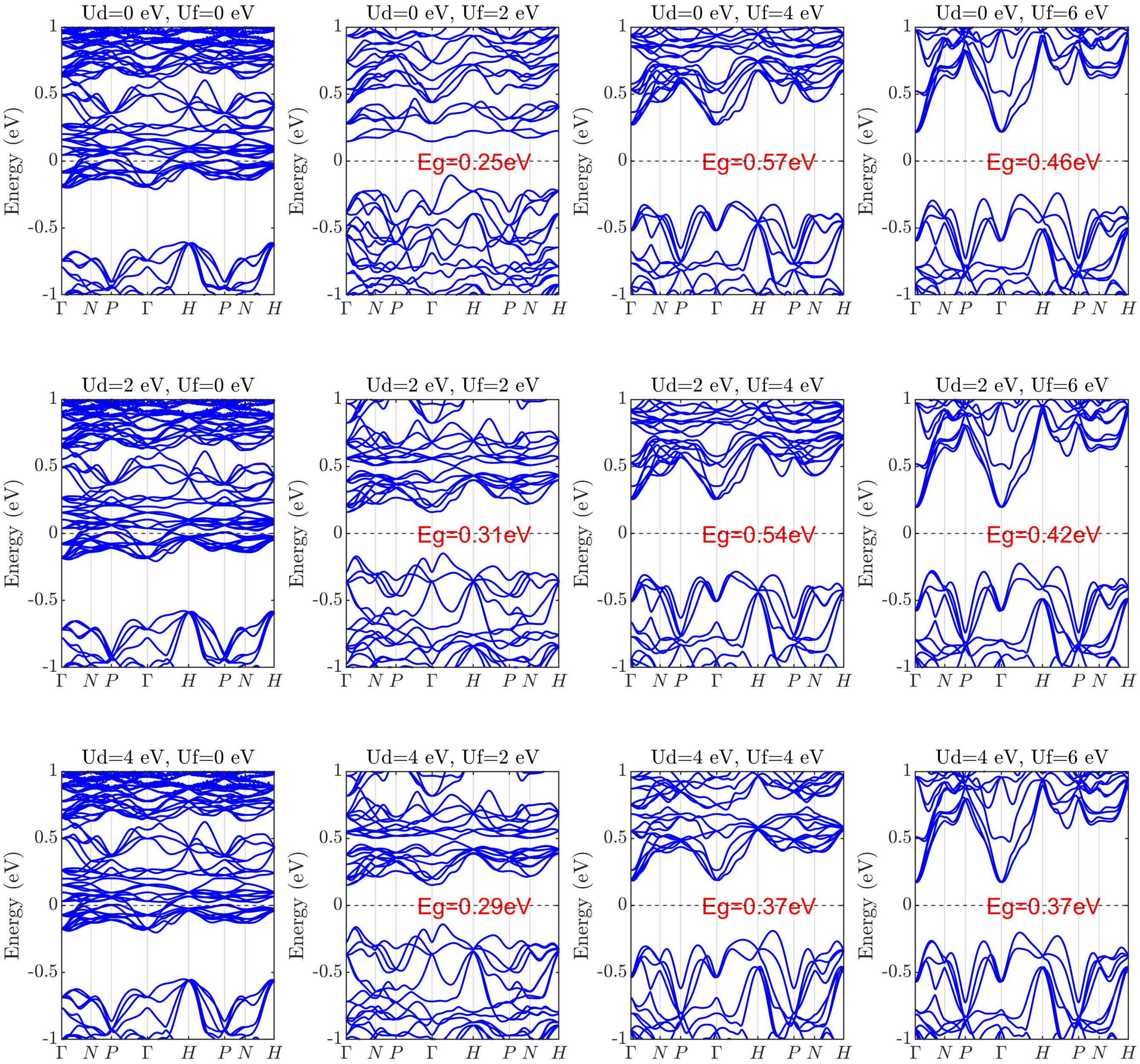

**Fig.** 4. Density Functional Theory (DFT) calculations, for a multitude of Hubbard *U* values applied to the Ni-*d* and U-*f* bands.